\documentclass[final]{svjour3}
\usepackage{gensymb}
\usepackage{graphicx,wrapfig,lipsum}
\usepackage{rotating}
\usepackage{amssymb}
\usepackage{amsmath}
\usepackage{mathptmx}
\usepackage{xcolor}
\usepackage{sidecap}
\usepackage{caption}
\usepackage{subcaption}
\usepackage{sidecap}
\usepackage{multirow}
\usepackage[symbol]{footmisc}

\usepackage{lineno}
\makeatletter
\journalname{Journal of Low Temperature Physics}

\begin{document}
\newcommand{\hdblarrow}{H\makebox[0.9ex][l]{$\downdownarrows$}-}
\titlerunning{Superconducting On-chip Fourier Transform Spectrometer}
\authorrunning{Basu Thakur, Steiger et al}
\title{Development of Superconducting On-chip Fourier Transform Spectrometers}

\author{R.~Basu Thakur$^{\mathcal{A},\mathcal{B}*}$\and A.~Steiger$^{\mathcal{C},\mathcal{D}}$\and S. Shu$^{\mathcal{A},\mathcal{B}}$\and F.~Faramarzi$^\mathcal{E}$ \and N. Klimovich$^{\mathcal{A},\mathcal{B}}$ \and P.K.~Day$^\mathcal{B}$\and E.~Shirokoff$^{\mathcal{C},\mathcal{D}}$\and P.D.~Mauskopf$^\mathcal{E}$  \and P.S.~Barry$^\mathcal{F}$ }

\institute{$\mathcal{A}\;$Dept. of Physics, California Institute of Technology, Pasadena, California 91125, USA
\\$\mathcal{B}\;$Jet Propulsion Laboratory (NASA), 4800 Oak Grove Dr., Pasadena, California 91109, USA
\\$\mathcal{C}\;$Kavli Institute for Cosmological Physics, 5640 S. Ellis Ave., Chicago, IL,60637, USA
\\$\mathcal{D}\;$Dept. of Astronomy \& Astrophysics, University of Chicago, 5640 S. Ellis Ave., Chicago, IL, 60637, USA
\\$\mathcal{E}\;$School of Earth \& Space Ex. \& Dept. of Physics, Arizona State University, Tempe, AZ 85281 USA
\\$\mathcal{F}\;$HEP Division, Argonne National Laboratory, 9700 South Cass Avenue., Argonne, IL,
60439, USA}

\maketitle

\begin{abstract}

Superconducting On-chip Fourier Transform Spectrometers (SOFTS) are broadband, ultra-compact and electronic interferometers. SOFTS will enable kilo-pixel spectro-imaging focal planes, enhancing sub-millimeter astrophysics and cosmology. Particular applications include cluster astrophysics, cosmic microwave background (CMB) science, and line intensity mapping. This article details the development, design and bench-marking of radio frequency (RF) on-chip architecture of SOFTS for Ka and W-bands. 

\keywords{Nonlinear Kinetic Inductance, Spectrometer, CMB, Line Intensity Mapping }

\end{abstract}

\section{Introduction}

In thin film superconductors like NbTiN and NbN, increasing supercurrent $I$ modifies the density of states, increasing kinetic inductance~\cite{Anthore:2003a}. This, in turn, alters the phase velocity, ultimately enabling current-controlled delays in a transmission line geometry. For a microstrip transmission line (length $\ell$, width $w$, inductance-per-square $L_{\Box}$ and impedance $Z_0$) with inductance and capacitance per unit length of $\mathcal{L}$ and $\mathcal{C}$ respectively, the current-controlled delay is given by Eq.~\ref{eq:1}. The characteristic currents $I_*, I_*'$ are determined by the material properties and device geometry. 
\begin{equation}
	\Delta \tau (I) = \ell\left(\sqrt{ \mathcal{L}(I) \mathcal{C}} - \sqrt{\mathcal{L}(I=0) \mathcal{C}}\right) \approx \frac{L_{\Box}\ell}{Z_0w} \left( \sqrt{[1+(I/I_*)^2 + (I/I_*')^4 ]} -1\right)
	\label{eq:1}
\end{equation}

We employ this core idea to realize Superconducting On-chip Fourier Transform Spectrometers (SOFTS~\cite{BasuThakur2020}). A broadband input is split in two parallel transmission lines where a relative phase delay is introduced with current biasing. And recombined signals form interferograms like a classical FTS. In this paper, we present a thorough superconducting circuit design, numerical characterization, device fabrication and calibration plans expanding on our previous work in 1-10 GHz and 25-40 GHz ranges~\cite{BasuThakur2020},\cite{Shibo1}.

\section{Mach-Zhender architecture}
 \vspace{-0.1 in}

\begin{figure}[h!]
     \centering
     \begin{subfigure}[b]{0.6\textwidth}
         \centering
         \includegraphics[width=\textwidth]{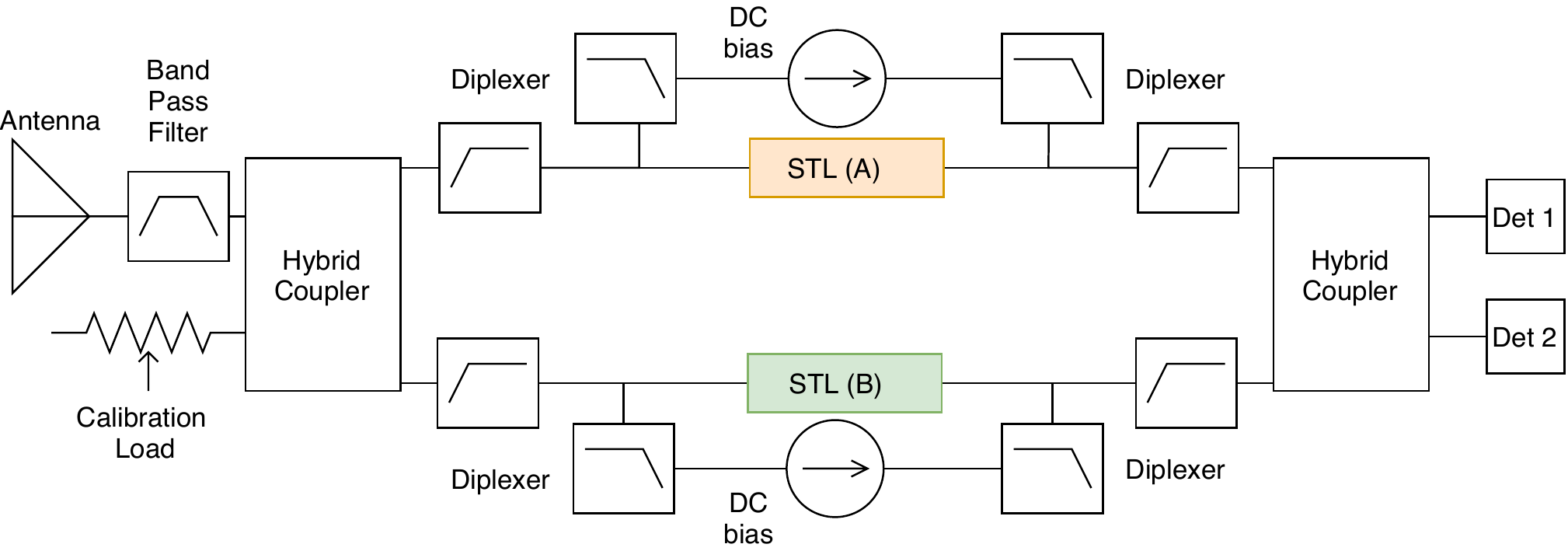}

     \end{subfigure}
   \hfill
     \begin{subfigure}[b]{0.33\textwidth}
         \centering
         \includegraphics[width=\textwidth]{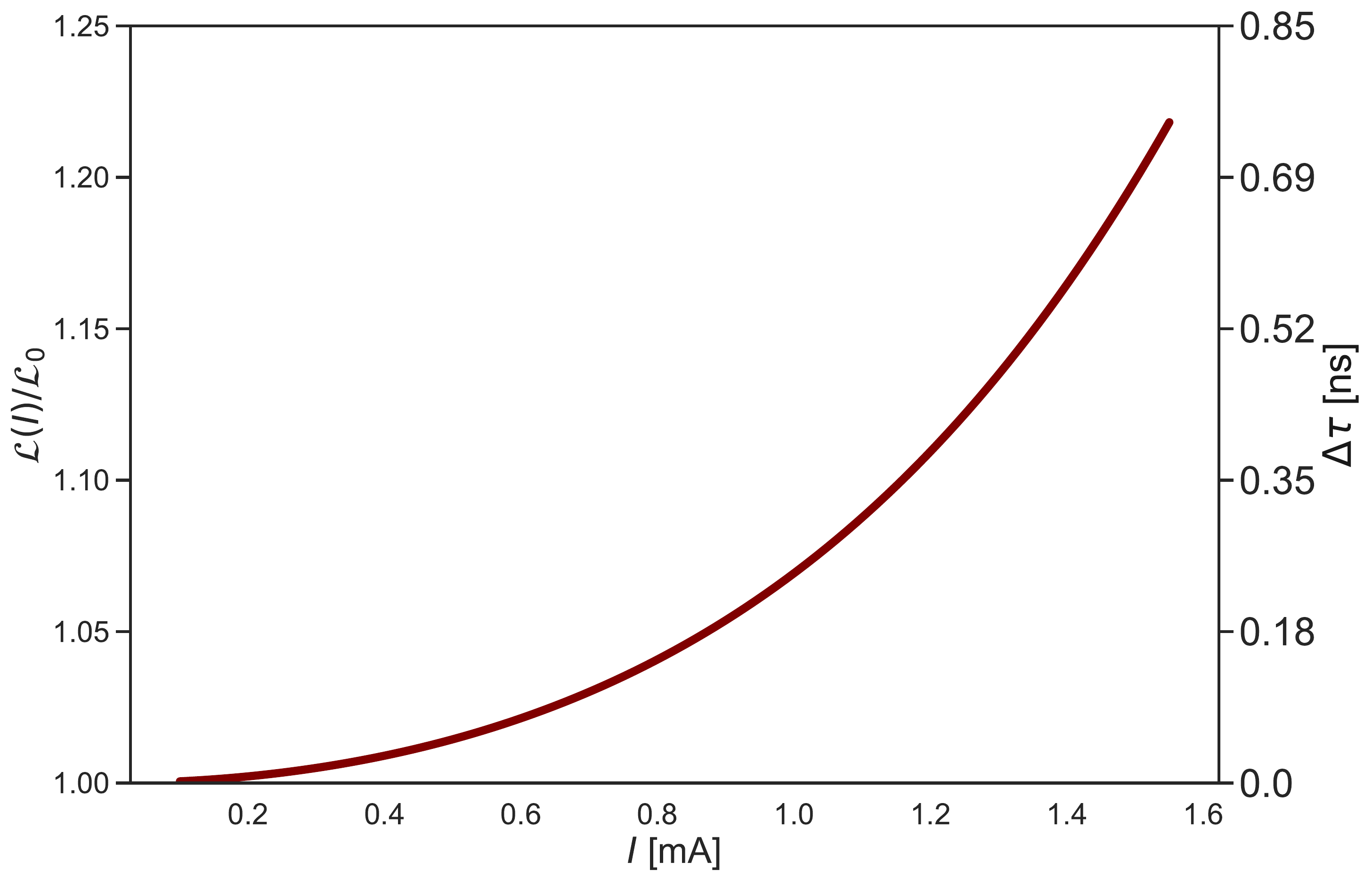}

     \end{subfigure}
        \caption{\small{(a) System level diagram for 4-port SOFTS (b) Delay with current derived from measurements~\cite{Shibo1}.} }
        \label{fig:softs_a}
\end{figure}

SOFTS is designed as a 4-port Mach-Zhender interferometer, Fig.~\ref{fig:softs_a}; Michelson and other architectures are also doable~\cite{PIXIE}. The two inputs are a broadband antenna with band-defining filter observing the sky and a bolometric load as a precision calibrator. These are combined with phasing via a hybrid coupler (HC). Superconducting Transmission Lines (STLs) connect to HC via diplexers. The diplexers DC bias the STLs. The STLs are effectively optical arms of a FTS. With DC current biasing one arm (potential audio-band AC biasing to be explored for multiplexing advantage), relative phase delay is added such that following the final HC, two detectors see the interfered power analogous to the symmetric and antisymmetric ports of a classical FTS. We ultimately Fourier transform the measured interferograms, where achievable frequency resolution is $\delta \nu = 1/\text{max}(\Delta \tau (I))$.  

\begin{figure}[h!]
     \centering
    \vspace{-30pt}
     \begin{subfigure}[b]{0.56\textwidth}
         \centering
         \includegraphics[width=\textwidth]{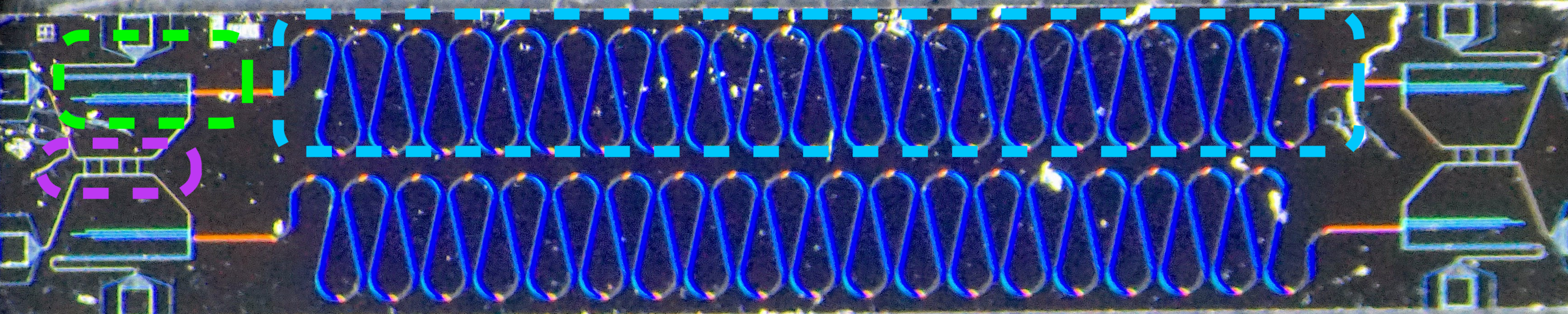}

     \end{subfigure}
     \hfill
      \begin{subfigure}[b]{0.33\textwidth}
     \centering
     \includegraphics[width=\textwidth]{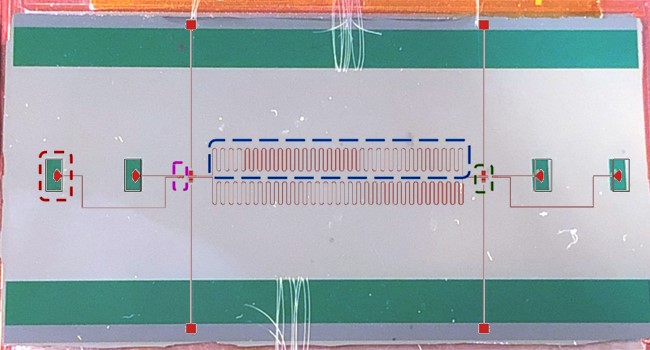}

 \end{subfigure}

\caption{\small{4-port SOFTS devices, colored dashed boxes denote sub-systems: \emph{Left}, Ka-band chip, hybrid coupler (purple), transmission line (blue), diplexer (green) \emph{Right}, W-band, with mask overlaid on photo: radial probe (red), hybrid coupler (purple), transmission line (blue) and bias tees (green). See table~\ref{tab:softs_tab1} for details.} }
\label{fig:softs_b}
\end{figure}
 \vspace{-25pt}
\begin{table}[h!]
\begin{tabular}{ |p{1cm}||p{2cm}|p{1cm}|p{1cm}|p{1.5cm}|p{1.2cm}|p{1.1cm}|  }
  \hline
  Band & chip-dimensions [mm] & Material & $T_c$ [K] & nonlinearity ($\Delta${L}$_{\textrm{max}}$/L) & $\delta \nu$ [GHz] & $\nu_{max}$[THz] \\
 \hline
 Ka   & 1.15$\times$6.08    & NbTiN & 14 & 20\% &   1.43~\cite{BasuThakur2020} & $\lesssim 1$\\
 W &   45 $\times$ 15  & NbN & 13 & 18 \% & 0.1 & $\lesssim 1$\\
 \hline
\end{tabular}
\caption{Parameters for two current SOFTS architectures.}
\label{tab:softs_tab1}
\end{table}

\begin{figure}[h!]
     \begin{subfigure}[t]{0.35\textwidth}
         \centering
         \includegraphics[width=\textwidth]{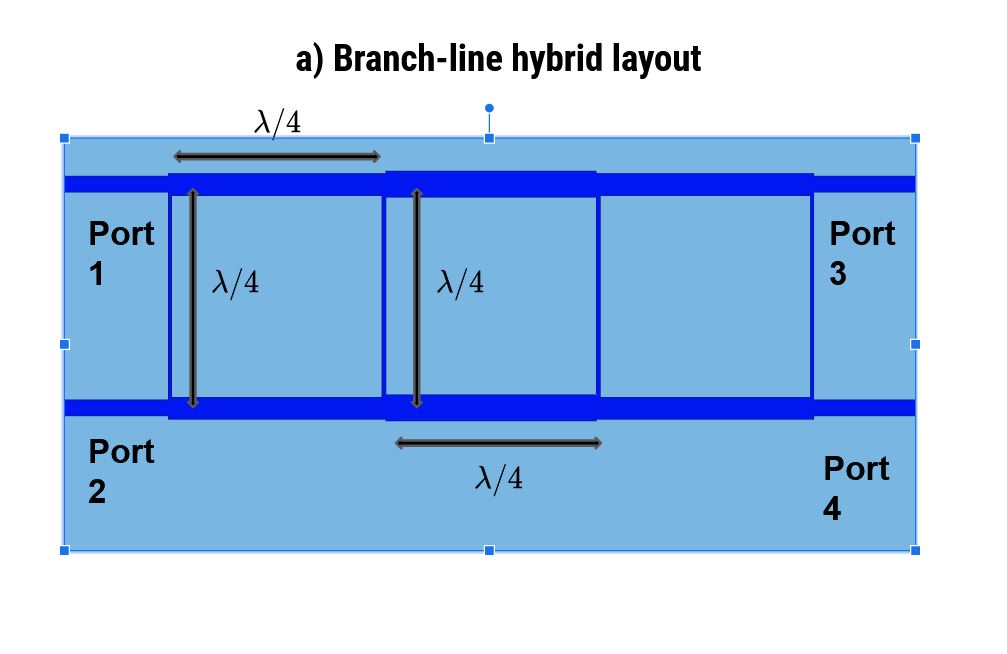}
         \label{fig:HC-W}
     \end{subfigure} 
     \centering
     \begin{subfigure}[t]{0.31\textwidth}
         \centering
         \includegraphics[width=\textwidth]{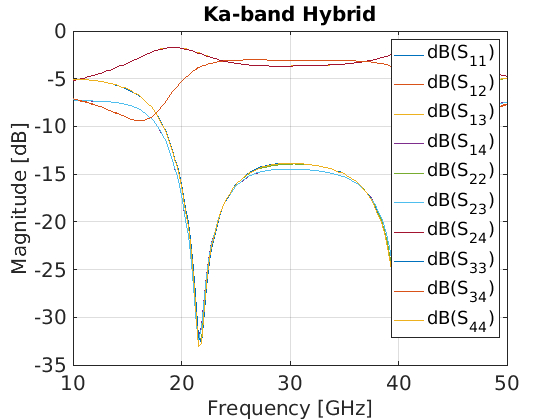}
         \put(-80,60){b)}
         \label{fig:HC-Ka}
     \end{subfigure}
     \begin{subfigure}[t]{0.31\textwidth}
         \centering
         \includegraphics[width=\textwidth]{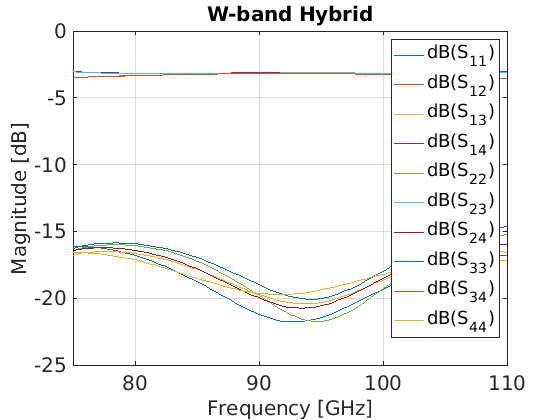}
         \put(-85,65){c)}
         \label{fig:HC-W}
     \end{subfigure}
     \vspace{-10pt}
        \caption{\small{ a) Superconducting Hybrid Coupler (HC) structure on-chip. We vary the $\lambda/4$ length-scale to operate in requisite bands. Simulated S-parameters for (b) Ka-band HC and (c) W-band HC  }}
        \label{fig:subsysKa1}
\end{figure}

\begin{figure}[h]
\vspace{-10pt}
         \begin{subfigure}[b]{0.6\textwidth}
         \centering
         \includegraphics[width=0.6\textwidth,height=0.25\textwidth]{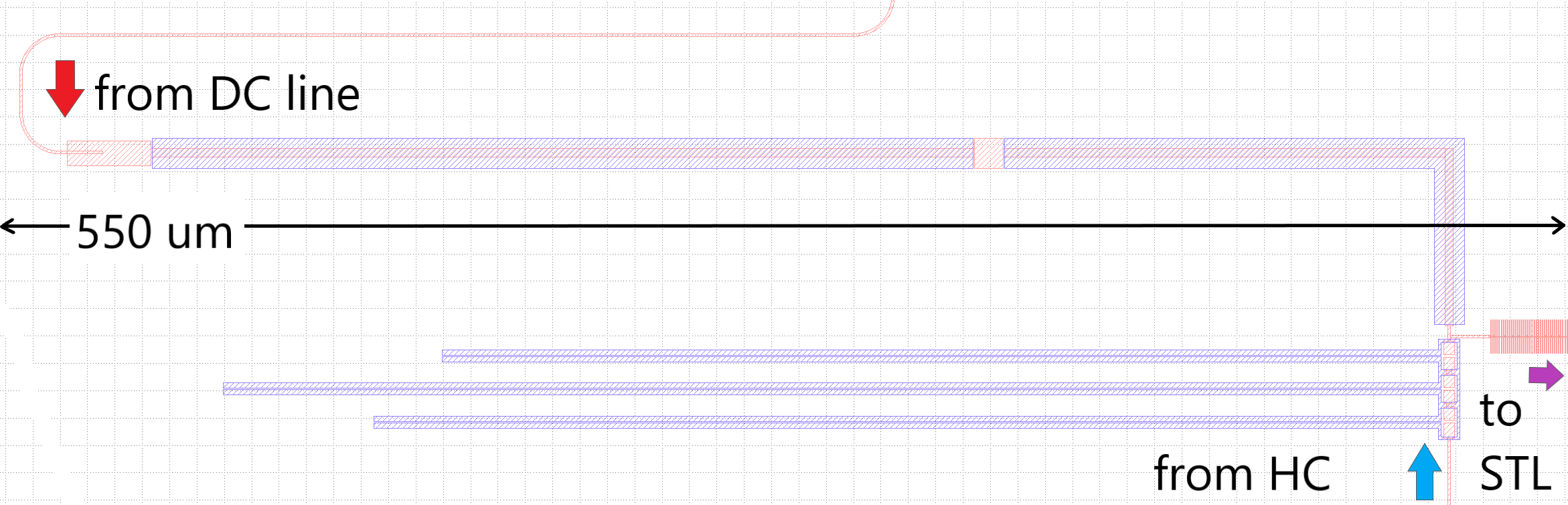}
         \put(-160,50){a)}
         \label{fig:DPX-Ka-mask}
     \end{subfigure}
         \hfill
         \begin{subfigure}[b]{0.34\textwidth}
         \centering
         \includegraphics[width=\textwidth,height=0.5\textwidth]{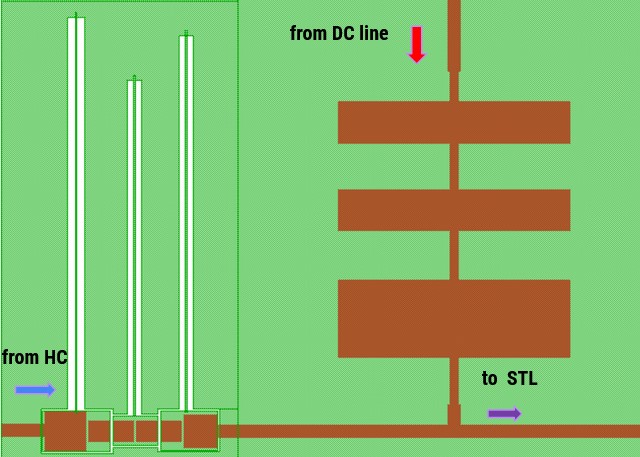}
         \put(-130,50){b)}
         \label{fig:DPX-W-mask}
     \end{subfigure}
        \vfill
            \begin{subfigure}[b]{0.33\textwidth}
            \hspace{10mm}
        \includegraphics[width=\textwidth,height=0.7\textwidth]{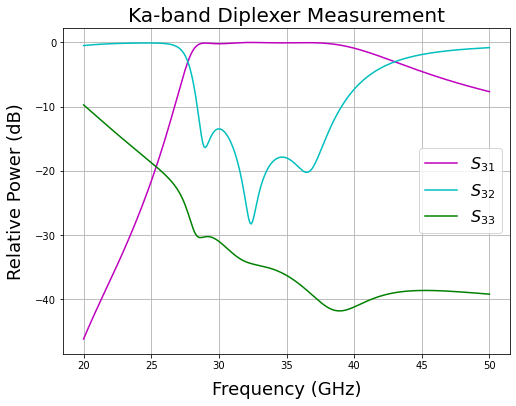}
         \put(-130,80){c)}
         \label{fig:DPX-Ka}
         \end{subfigure}
         \hfill
     \begin{subfigure}[b]{0.33\textwidth}
         \centering
         \includegraphics[width=\textwidth,height=0.7\textwidth]{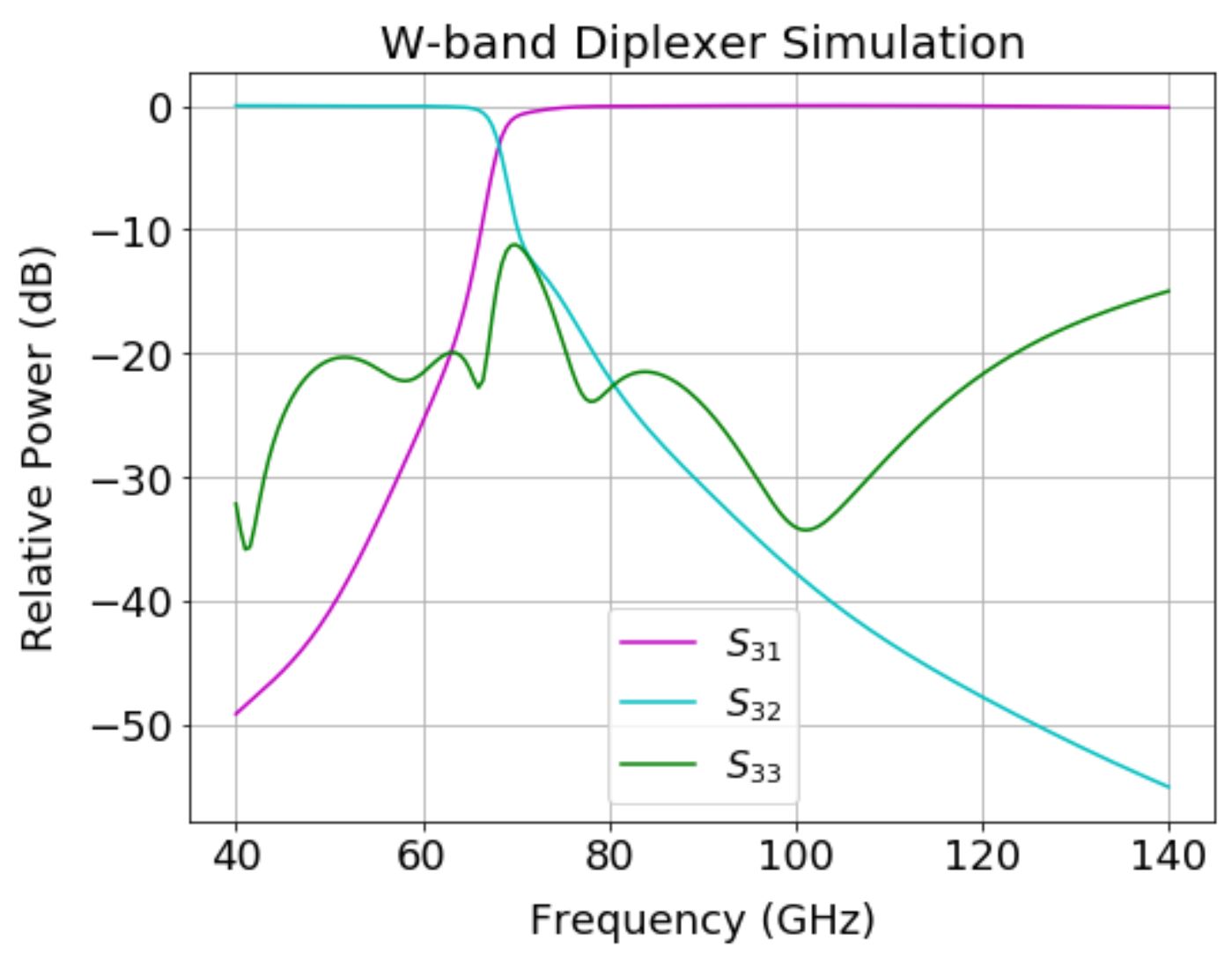}
         \put(-130,80){d)}
         \label{fig:DPX-W}
         \end{subfigure}
   
        \caption{\small{ On-chip diplexers realized with capacitative stubs and inductive lines: (a) Ka-band (b) W-band (c) Ka-band diplexer simulated scattering parameters (d) W-band diplexer simulated scattering parameters. }}
        \label{fig:subsysKa2}
\end{figure}
\vspace{-3 mm}
We are testing two SOFTS architectures in parallel with Ka and W-band devices, see Table~\ref{tab:softs_tab1} and Fig.~\ref{fig:softs_b}. The variations in superconducting material, geometry, and optical coupling are intentional. The W-band chip has a larger footprint than the Ka-band device because (i) for a 4-port split block, we needed the probes in a row which takes up more space (ii) to lower $\delta \nu$ (frequency resolution) we use a long STL for the W-band device. The Ka-band chip in contrast is optimized for lower resolution and high compactness. We will take the best aspects of each design and develop a unified SOFTS architecture for particular science cases. For CMB-science we need $\delta \nu \sim O(1)$ GHz, and for line-intensity mapping $\delta \nu \sim O(0.1)$ GHz. SOFTS resolution is tunable, i.e., we can always increase $\delta \nu$ by lowering the bias current. Here we quote the smallest $\delta \nu$ which depends on device geometry and material. Fig.~\ref{fig:subsysKa1} and \ref{fig:subsysKa2} show design and simulation of on-chip HCs and diplexers used in our devices.
\section{RF Analysis}
\vspace{-3 mm} 
\subsection{RF cascade simulation}
\vspace{-3 mm} 
\begin{SCfigure}
     \caption{\small{SOFTS cascade-network, `a's /`b's denote incident/ reflected voltages, super-scripts and sub-scripts denotes individual subsystems and port indices. Diplexers suppressed for simplicity}}
    \includegraphics[width = 0.6\textwidth]{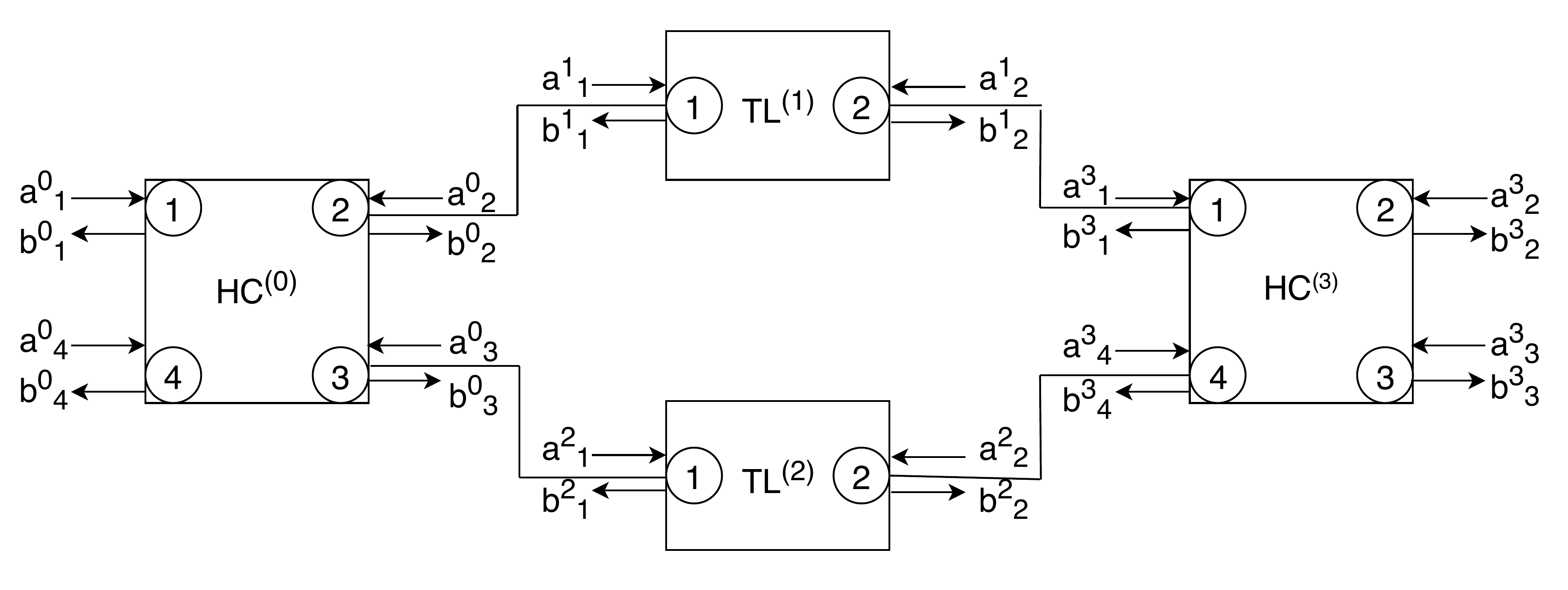}
    \label{fig:aa_circuit}
\end{SCfigure}
N-port subsystem S-matrices are linked as shown in Fig.~\ref{fig:softs_a} and cascade simulations are run with MathWorks' RF Blockset software\footnote{https://www.mathworks.com/help/simrf/}. We input single tones on port 1 of first hybrid-coupler  (HC$^{\left(0\right)}$), monitoring the output power (ports 2 and 3 of HC$^{\left(3\right)}$)  and reflected power in the second input port (port 4 of HC$^{\left(0\right)}$), Fig.~\ref{fig:aa_circuit}. All powers are monitored as dissipation across 50 $\Omega$.In the ideal case, i.e., for negligible reflections and loss from the transmission lines, and low cross-talk between output ports of each hybrid-coupler, we can model the power measured in the symmetric and anti-symmetric ports (ports 2 and 3 of HC$^{\left(3\right)}$). For unit voltage input on the antenna port, this is given by Eq.~\ref{eq:power_meas_ideal}, where $\Delta\tau=\tau_{2}-\tau_{1}$. 
\small
\begin{equation}\label{eq:power_meas_ideal}
\left.\phantom{}
\begin{split}
 P^{(3)}_2 & = \frac{1}{50 \Omega} \left|S^{(3)}_{21}(\nu)S^{(0)}_{21}(\nu) + e^{-i2\pi\nu \Delta \tau(I)}S^{(3)}_{24}(\nu)S^{(0)}_{31}(\nu) \right|^2 \\
P^{(3)}_3 & = \frac{1}{50 \Omega} \left|S^{(3)}_{31}(\nu)S^{(0)}_{21}(\nu) + e^{-i2\pi\nu \Delta \tau(I)}S^{(3)}_{34}(\nu)S^{(0)}_{31}(\nu) \right|^2 
\end{split}
\right\}
\Rightarrow
P^{(3)}_{2/3}\approx\frac{1\text{V}^2}{50 \Omega \cdot 2} \left(1 \pm \cos (2\pi \nu \Delta \tau (I))\right)
\end{equation}
\normalsize
For the 90-deg hybrid-coupler, $S_{34} \approx S_{21} \approx i$ and $S_{31} \approx S_{24} \approx 1 $, and eq.~\ref{eq:power_meas_ideal} reduces to the standard FTS results of cosine modulations as shown above. 
RF cascade simulations allow us to comprehensively model frequency dependencies, multi-path effects from reflections and cross-talk, from which we can expect mild anharmonicities. The ability to comprehend these anharmonicities is indeed a major advantage for SOFTS, i.e., we can understand the spectrometer performance as a pure circuit model, as compared to optical FTS where multipath effects are challenging to accurately model and correct for. Here we are using simulated S-parameters; measured $\textbf{A}_{\text{SOFTS}}$, e.g., our previous publication~\cite{BasuThakur2020},  can also be used.\\

\begin{figure}[h!]
     \centering
     \vspace{0 mm}
     \begin{subfigure}[b]{0.4\textwidth}
         \centering
         \includegraphics[width=\textwidth]{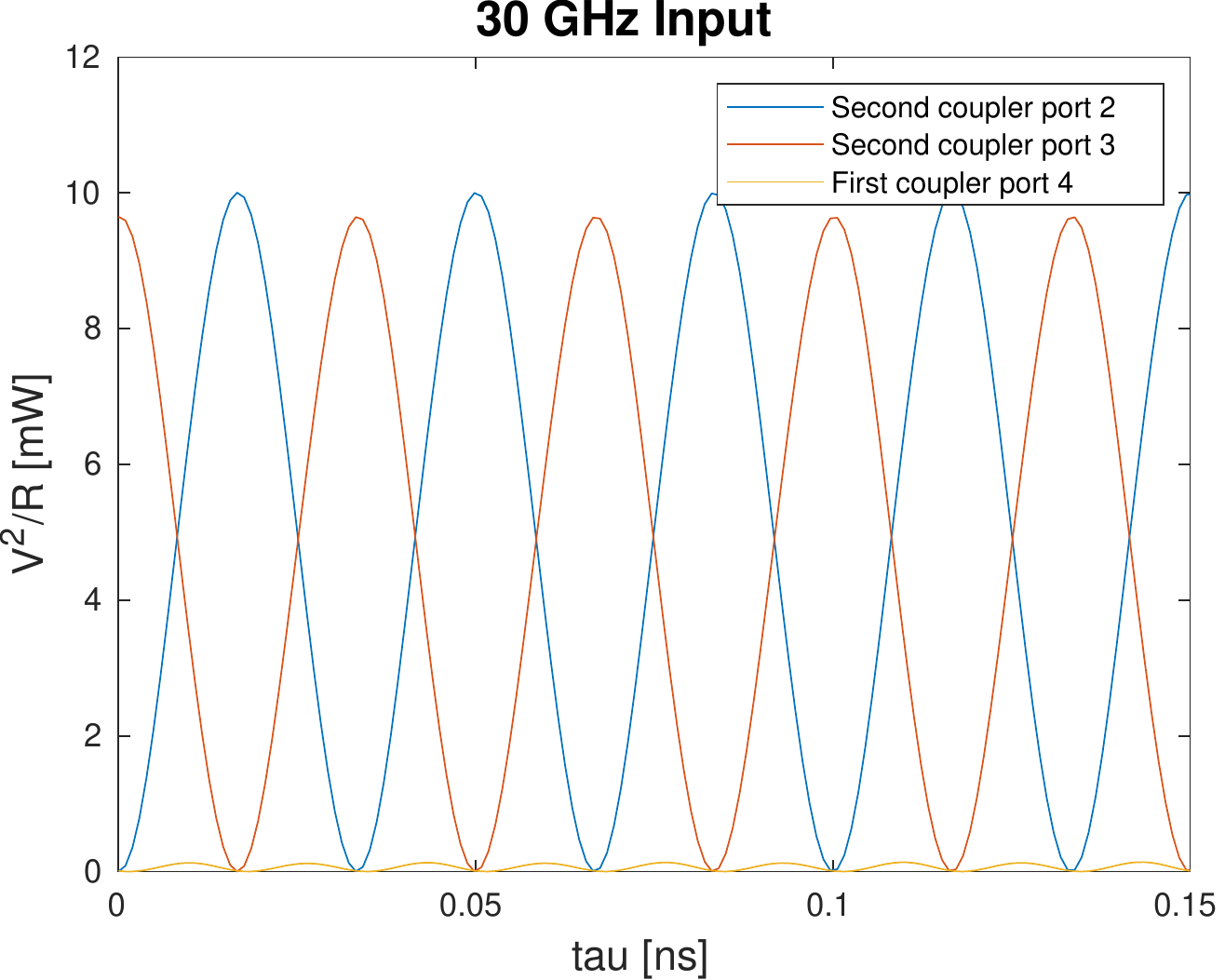}
         \put(-125,100){a)}
         \label{fig:nonideal_interfs}
     \end{subfigure}
     $\quad$
     \begin{subfigure}[b]{0.42\textwidth}
         \centering
         \includegraphics[width=\textwidth]{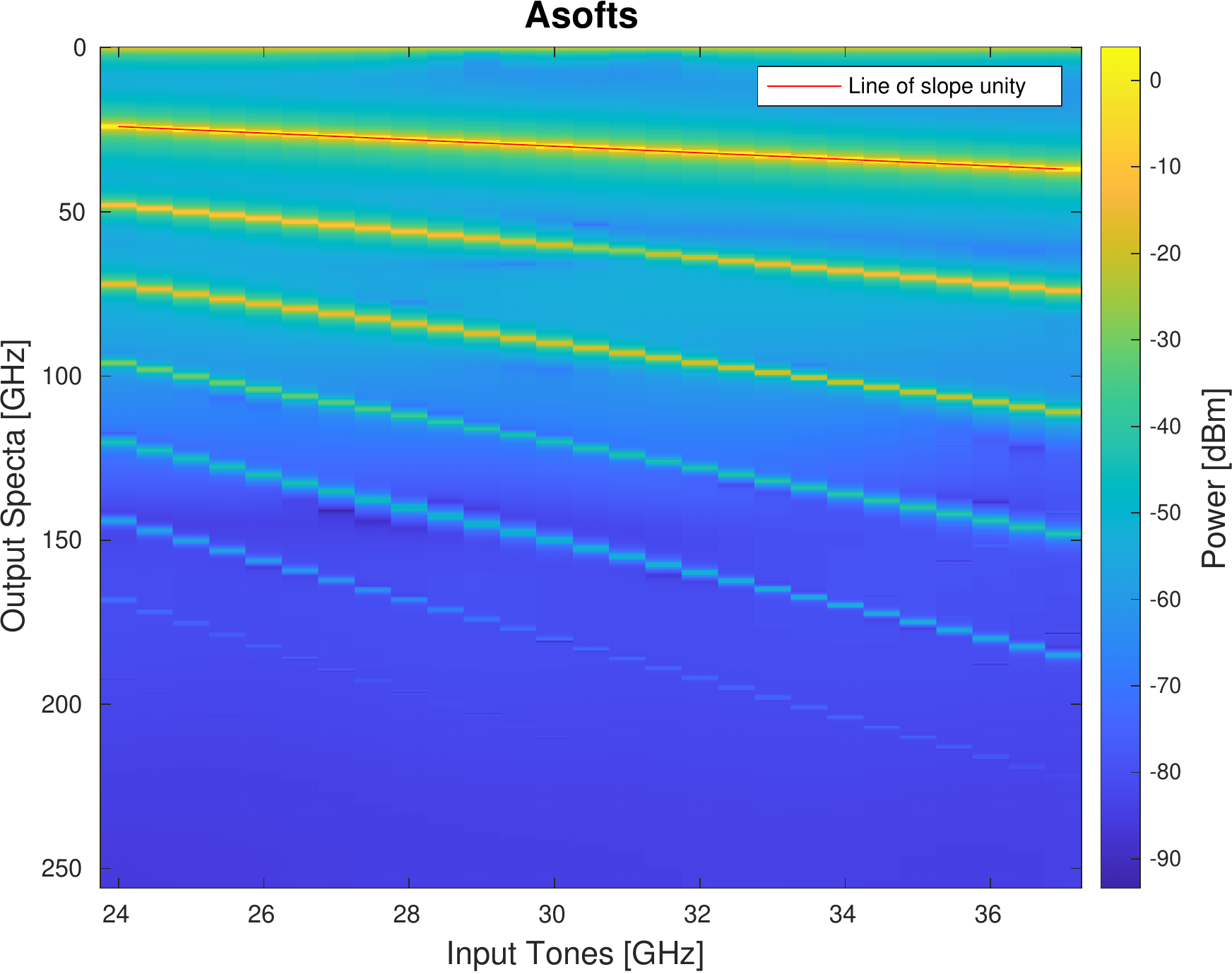}
         \put(-125,98){b)}
         \label{fig:ASOFTS}
     \end{subfigure}
        \caption{\small{RF simulations of SOFTS in Ka-band: (a) Interferograms (zoomed in) for a 30 GHz input tone (b) $\textbf{A}_{\text{SOFTS}}$ (transfer function matrix) for port-2 of the output hybrid coupler.}}
        \label{fig:makingASOFTS}
\end{figure}

From Eq.~\ref{eq:1} and prior work~\cite{Shibo1},\cite{BasuThakur2020} we expect up to 2 ns of delay. For every single-tone input, we scan over this range to produce interferograms and their FFTs generate a transfer function ($\textbf{A}_{\text{SOFTS}}$), i.e., observed frequencies for single-tone inputs. Fig.~\ref{fig:makingASOFTS} shows relevant figures for Ka-band studies and W-band simulations are done identically. Digital signal processing with the transfer function enables accurate spectral recovery.

\vspace{-5 mm} 
\subsection{Error correction}
\begin{figure}[h!]
     \centering
     \begin{subfigure}[t]{0.37\textwidth}
         \centering
         \includegraphics[width=\textwidth]{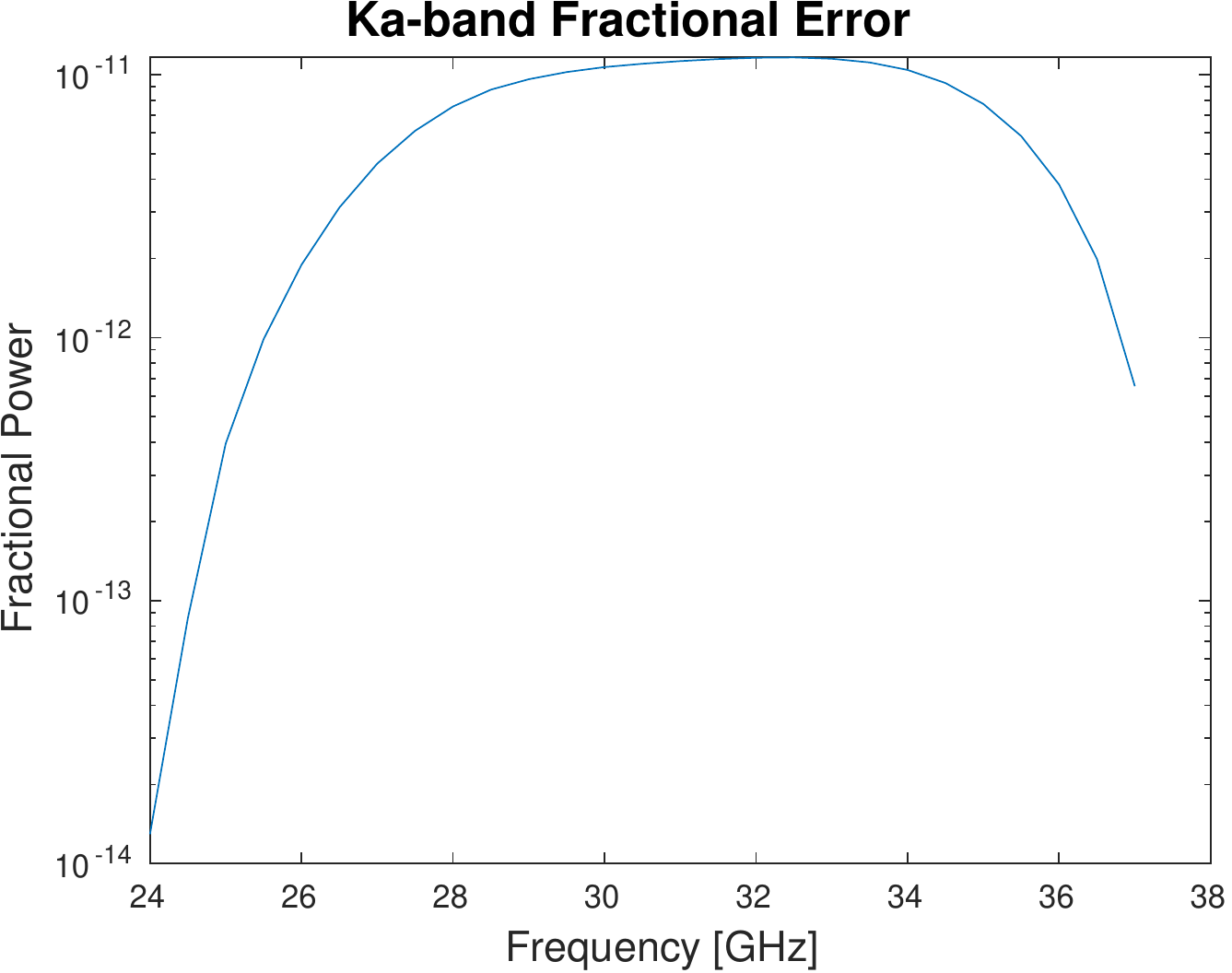}
         \label{fig:Ka Fe}
     \end{subfigure}
     $\quad$
         \begin{subfigure}[t]{0.37\textwidth}
         \centering
         \includegraphics[width=\textwidth]{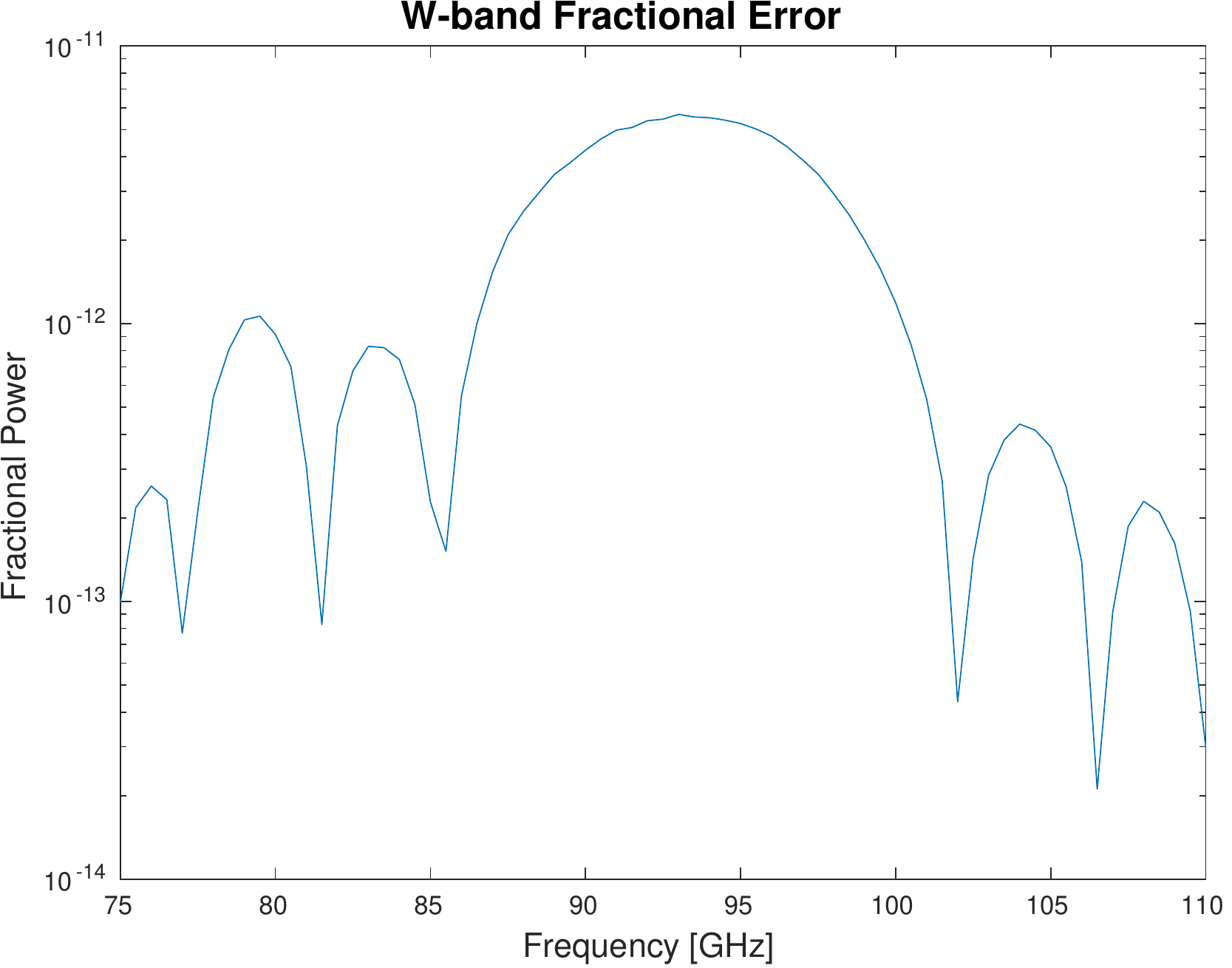}
         \label{fig:W Fe}
     \end{subfigure}
     \caption{\small{Ka-band and W-band fractional errors following complete RF cascade simulations. }}
     \label{fig:fe}
\vspace{-2mm}
\end{figure}

\vspace{-3 mm} 
Error correction implies accounting for device non-ideality generated anharmonicities so that we have accurate spectral recovery. Each input tone is a unit vector in frequency space, named $\Vec{U}^k$, where \emph{only} the $k^{th}$ element is 1, e.g. $\nu_{min} =$
$
\begin{bmatrix}
1, & 0, & 0, ...
\end{bmatrix}^{T}
$, $\nu_{min}+\delta\nu =$
$
\begin{bmatrix}
0, & 1, & 0, ...
\end{bmatrix}^{T}
$. Each single tone input generates multi-tone output given by $\Vec{V}^k = \textbf{A}_{\text{SOFTS}}\cdot\Vec{U}^k$, see Fig.~\ref{fig:makingASOFTS}. Suppose that $B_{true}$ is the true multichroic sky-signal and $B_{obs}$ is the SOFTS spectrum that is readout. Since $\Vec{U}^k$ is essentially a delta function in frequency space, we pursue inversion following least-squares method, Eq.~\ref{eq:Green's}.
\begin{equation}\label{eq:Green's}
  \left(\textbf{A}_{\text{SOFTS}}^{T}\textbf{A}_{\text{SOFTS}}\right
   )^{-1}\textbf{A}_{\text{SOFTS}}^{T}\Vec{V}^k = \delta_{\nu,\nu_k}\Rightarrow
    B_{true} = \left(\textbf{A}_{\text{SOFTS}}^{T}\textbf{A}_{\text{SOFTS}}\right
   )^{-1}\textbf{A}_{\text{SOFTS}}^{T}\cdot B_{obs}
\end{equation}
$\ \left(\textbf{A}_{\text{SOFTS}}^{T}\textbf{A}_{\text{SOFTS}}\right
   )^{-1}\textbf{A}_{\text{SOFTS}}^{T}$ is analogous to a Green's function for SOFTS devices. We demonstrate spectral reconstruction with fractional errors $ | B_{obs}-B_{true}|/B_{true}\lesssim 10^{-11}$ by considering the CMB spectrum as measured by SOFTS, Fig.~\ref{fig:fe}. This is a major achievement over classical optical FTSs where reconstruction of multi-path and alignment issues are far less accurate~\cite{optical_FTS}. Our error is fundamentally set by circuit non-idealities across the chip, caused by impedance mismatches originating from practical limits of fabrication. Here we demonstrate spectral recovery without noise. Robust recovery with noise is also possible in this framework~\cite{Halko}, and has been demonstrated for photon and detector noise~\cite{Steiger}.
\vspace{-7mm}
\section{Optical coupling and device hardware}
\begin{figure}[h!]
     \centering
     \begin{subfigure}[t]{0.29 \textwidth}
         \centering
         \includegraphics[width=\textwidth]{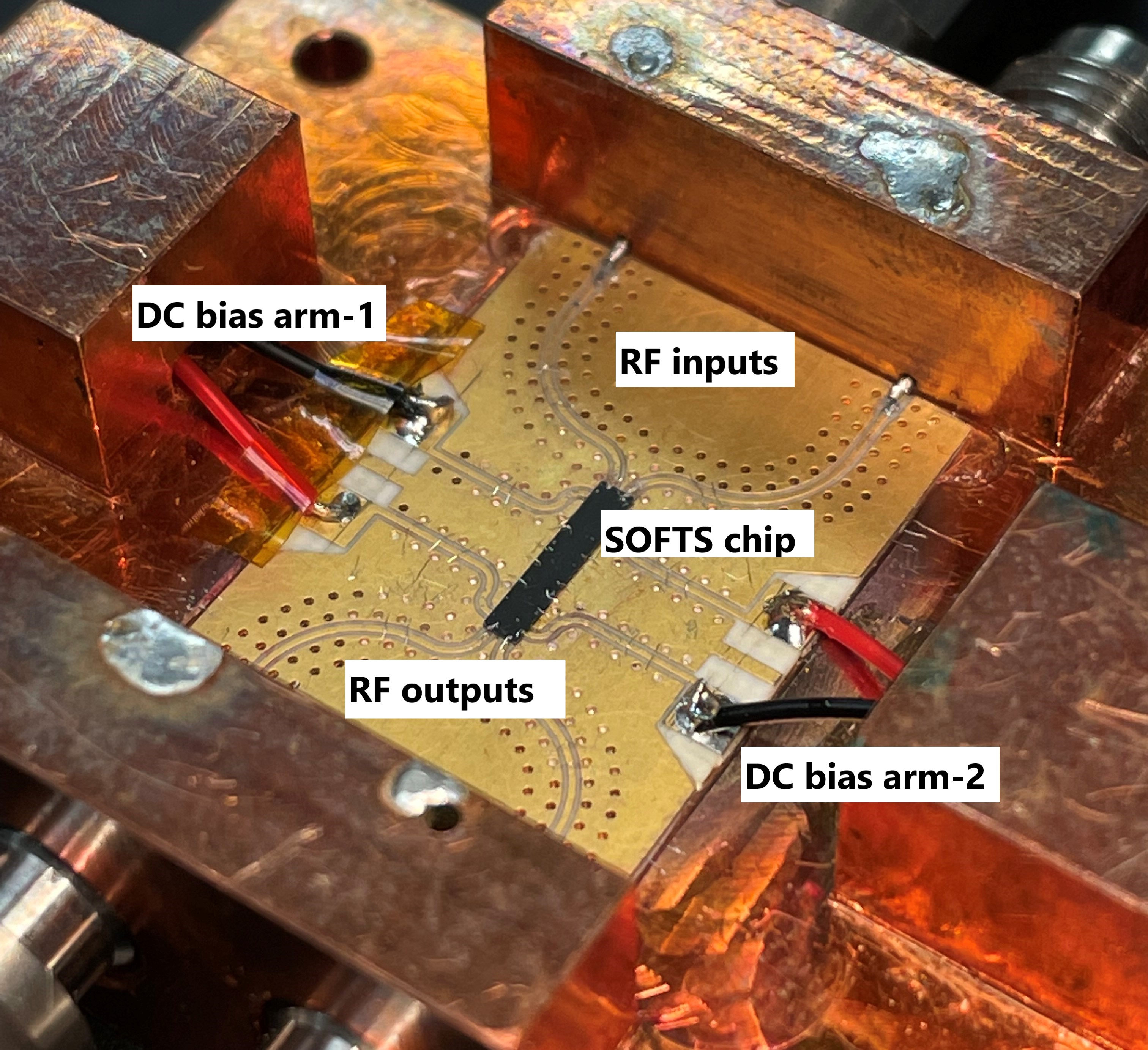} \put(-100,80){\textcolor{white}{a)}}
         
     \end{subfigure}
     \begin{subfigure}[t]{0.43\textwidth}
         \centering
         \includegraphics[width=\textwidth]{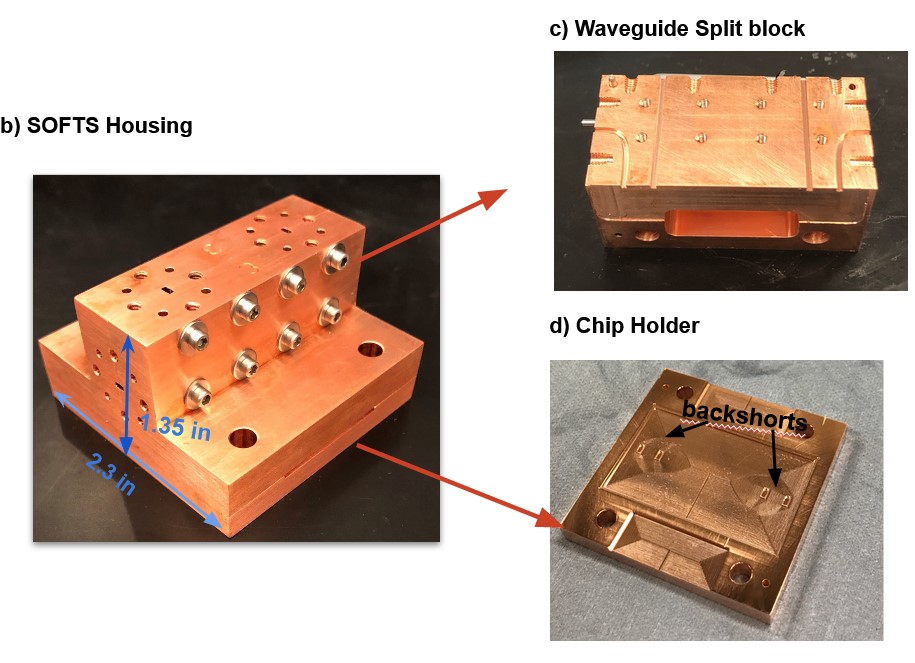}
         
     \end{subfigure}
        \caption{\small{a)The Ka-band SOFTS chip sits in the PCB cut-out, and is wirebonded to the leads. PCB routes the signals to DC and RF connectors at the edge of the copper module. b) Housing assembly for W-band SOFTS chip includes four waveguide flange fittings for each port and through holes for mounting. d) A split block is used to transition from the waveguides to the W-band on-chip probes. d) The chip holder design includes backshorts for the probes and trenches for the DC bias PCBs.  }}
        \label{fig:KaWhouse}
\end{figure}

\textbf{Ka-band SOFTS} in essence is an inverted microstrip architecture where 35 nm thick and 250 nm wide NbTiN is the workhorse superconductor,  Fig.~\ref{fig:softs_b}. The STL fabrication is a stepper driven process and all fabrication steps are identical to our published work on the measurement of Ka-band phase-delay~\cite{Shibo1}. The minimum resolved frequency for this device is expected to be 1.43 GHz, based on prior measurements~\cite{Shibo1}. The STL design is mostly band-independent and the maximum frequency is limited by the $\sim$1.2 THz gap for NbTin~\cite{klimovich}. It will be slightly reduced during operation due to the change in density of states from the applied bias current. For operation near $I_c$ the fractional reduction in the gap $\Delta(\Gamma)/\Delta_0 \approx 0.9$~\cite{Anthore:2003a}, still allowing for maximum frequencies $\sim$ 1 THz.
We have fabricated a printed circuit board (PCB) for mounting the SOFTS chip, and an OFHC copper housing to encase the chip and PCB for laboratory testing, Fig.~\ref{fig:KaWhouse} (a). We simulated the PCB over the Ka-band and measured $<$ -20 dB cross-talk and $<$-10 dB reflections. \emph{Ultimately for antenna coupled SOFTS the PCB is unnecessary.} The actual SOFTS chip is $\sim$ 6 mm $\times$ 1 mm. The PCB is necessary to make RF and DC wire bonding connections to the device and connect to the device housing ports. \\


\textbf{W-band SOFTS} housing with waveguide coupling has been designed and fabricated and coupling is done using a radial probe ~\cite{Farzad}. The STLs were formed from a 40 nm thick deposited niobium nitride film etched using a reactive ion etching (RIE) process. Other circuit elements, including the probes and the hybrids, were formed from a 150 nm thick niobium (Nb) film using a liftoff process. Silicon nitride was used as the dielectric layer ($500$ nm) and it was deposited on top of the circuit using a Plasma-Enhanced Chemical Vapor Deposition (PECVD) method. The ``skyplane" was then deposited on top of the dielectric. Lastly, the silicon below the probes is etched away for improved probe coupling. Chip-housing consists of three parts, with the top parts being the split-block waveguide as shown in Fig.\ref{fig:KaWhouse} (b-d), and a chip holder, which consists of the waveguide backshorts and acts as a heat sink.  We have two extrude cuts on the chip holder for DC-biasing circuit boards.
\vspace{-1cm}
\vspace{-1cm}
\section{Configuration and application examples}

While specific science requirements will define SOFTS-based kilopixel array architectures~\cite{Steiger}, we consider SOFTS devices covering 90-270 GHz and compare\footnote{This is not a comparison of scientific capabilities or sensitivities. We intended to outline SOFTS focal plane architecture in the context of the current kilopixel arrays.} with SPT-3G, a state-of-the-art CMB telescope~\cite{SPT3G_AA}. Frequency resolution determines STL length, the largest element in a SOFTS chip. For $\delta \nu \approx 4$ GHz, a sufficient resolution for CMB science cases, SOFTS will occupy $\approx2$ mm$^2$ as scaled from the Ka-band device. A 90-270 GHz antenna is $\approx10$ mm$^2$ and dominates focal plane area occupation. Therefore we can commensurately fit $\approx2100$ SOFTS pixels compared to 2690 in SPT-3G. Each SPT-3G pixel has 6 bolometers (3 coarse spectral bands $\times$ 2 polarizations) and SOFTS will have 4 bolometers per pixel (2 sum/difference ports $\times$ 2 polarizations); 33\% reduction in readout burden. Instead of 3 coarse bands in SPT-3G, SOFTS allows $(270-90)~\text{GHz}/4~\text{GHz} = 45$ spectral channels. Reducing DC bias lowers channel counts (commensurately increases sensitivity), should dynamic optimization be needed. On a focal plane similar to state-of-the-art, we will therefore have 1/3$^{rd}$ fewer detectors and $\times 15$ more spectral channels, all with similar pixel counts. Thus SOFTS enables kilopixel spectro-imaging in the sub-millimeter. Biasing $O(10^3)$ devices is non-trivial, though there is precedent of time-division multiplexing~\cite{BA_LM}. We are exploring AC biasing ($\sim$10 kHz) which can enable frequency domain multiplexing~\cite{SPT3G_AA}. \\

High accuracy CMB spectral distortions (CMB-SD) is an emerging field that needs new technologies~\cite{Super_PIXIE}. SOFTS can be introduced between the antenna and detector, e.g. Fig.~\ref{fig:softs_a}, whilst maintaining the general architecture of kilopixel arrays as discussed above. Such multiple simultaneous spectroscopic ``eyes-on-the-sky'' enables measuring CMB-SD. The SPT-3G comparative design above, which is \emph{not optimized} for CMB-SD yields sensitivities of $\sim 10$ Jy/sr, approaching other optimized mission concepts~\cite{Super_PIXIE}. Similar sensitivity will allow SOFTS designed for THz operation\footnote{New materials such as MgB$_2$ are under R\&D to explore SOFTS operations in the THz, as the superconductors discussed here may not be optimal.} to perform line intensity mapping (LIM) studies. Observational techniques between these fields overlap significantly and as discussed in literature~\cite{Kovetz2019,Serra}. While CMB Spectral distortion will probe very early universe physics, LIM will probe reionization physics and structure formation.

\section{Conclusion}
We have outlined detailed circuit modeling of Superconducting On-chip Fourier Transform Spectrometers (SOFTS) and discussed device design and hardware progress for Ka and W-band SOFTS, including the fabrication layout of SOFTS chips and their necessary optical coupling technologies. These bands were chosen based on the commercial availability of VNA and parts. However, we can rescale the hybrid and diplexer elements and retune the transmission line impedance, therefore SOFTS design is largely frequency band independent. Ultimately we intend for antenna and detector coupled SOFTS ~\cite{Steiger}. Furthermore we elucidated comprehensive RF cascade simulations of our complete devices, and demonstrated that such on-chip circuits have fractional errors in spectral recovery at levels of $\lesssim 10^{-11}$. Although due to COVID-19 device fabrication and testing has been delayed, our imminent work will involve measurements of both Ka and W-band devices. 
\begin{acknowledgements}
We thank Robert Webber (Caltech) for pointing  out noise spectral construction is possible with our framework~\cite{Halko}. We thank Rick LeDuc at JPL-MDL for device fabrication support. The W-band housing is being made using micro-mill machining by Matt Underhill at ASU. Undergraduate students C. Bell, E. Linden and E. Rapaport assisted with hardware assembly and participated through JPL/Caltech SURF. This research is supported by the NASA award NNH18ZDA001N-APRA, and by the University of Chicago College Summer Research Scholarship. 
\end{acknowledgements}
 \vspace{-0.3 in}
\bibliographystyle{IEEEtran}
\bibliography{LTDlib2021.bib}


\end{document}